# NIZKCTF: A Non-Interactive Zero-Knowledge Capture the Flag Platform

**Paulo Matias,** *Federal University of São Carlos, Brazil*

**Pedro Barbosa,** *Federal University of Campina Grande, Brazil*

**Thiago N. C. Cardoso,** *Hekima, Brazil*

**Diego M. Campos,** *University of São Paulo, Brazil*

**Diego F. Aranha,** *University of Campinas, Brazil*

*Capture the Flag (CTF) competitions are increasingly important for the Brazilian cybersecurity community as education and professional tools. Unfortunately, CTF platforms may suffer from security issues, giving an unfair advantage to competitors. To mitigate this, we propose NIZKCTF, the first open-audit CTF platform based on non-interactive zero-knowledge proofs.*

*Educational technology, Cryptography, Distributed information systems*

A critical element of a robust cybersecurity strategy is developing talent in modern technological security issues. Research shows that the United States is the most prepared country against cyber attacks,[1] however, there is also a problem of quantity and quality of professionals, especially when it comes to more sophisticated skills such as security by design, defensive programming, applied cryptography, threat intelligence and forensic analysis after a compromise.[2] This problem is aggravated in developing countries, where access to bleeding-edge resources for professional training based on real-world experience is limited.

In order to reduce the shortage of cybersecurity professionals, companies, schools, universities and military institutions have been promoting Capture the Flag (CTF) competitions around the world to foster the engagement of professionals in cybersecurity topics. CTF competitions are usually designed to serve an educational purpose to give participants experience with computer security problems from a wide spectrum of technical areas, as well as conducting and reacting to the sort of attacks found in the real world. CTF competitions can also serve as a convenient recruiting tool to fill specific positions with highly-skilled talent. Reverse-engineering, exploitation, forensics, web programming and cryptanalysis are among the typical required skills in CTF competitions. Because CTF competitions are inexpensive to organize and run, they are strategic to countries such as Brazil, allowing the local community to interact and compete with international players.

There are two styles of CTF competitions: attack/defense and jeopardy. In an attack/defense competition, each team is given a machine to defend on an isolated network. Teams are scored based on both their success in defending their assigned machine and on their success in attacking the other teams' machines. Jeopardy competitions are more common and usually involve multiple categories of problems, each of which contains a variety of questions of different values and levels of difficulty. A correct solution to a problem reveals a *flag*, which is submitted to the scoring platform for points. Teams attempt to earn the most points in the competition's time frame (e.g., 24 hours), but usually do not directly attack each other. Rather than a race, this style of gameplay encourages taking time to approach challenges and prioritizes quantity of correct submissions over the timing.

An important concern is to protect the CTF software platform against attacks. The platform is usually responsible for storing the flags and updating the scoreboard as the contest progresses. Unfortunately, because CTF platforms suffer from the same software security issues as any software component and due to incentives from the high competitiveness in such environments, it is common to find teams targeting the platform instead of the challenges. There are no independently verifiable guarantees that the teams really solved the challenges and the scoreboard is correct. Successful attacks against the platform arguably demonstrate relevant skills, but organizers may be more interested in enforcing the rules and rewarding solutions for the challenges, due to sponsorship duties or focused recruiting efforts.

In this paper, we propose a novel platform for jeopardy-style competitions called NIZKCTF: Non-Interactive Zero-Knowledge Capture the Flag Platform. With NIZKCTF, there are no flags stored in the platform, and therefore, for a solved challenge, the team does not submit the flag itself, but a public zero-knowledge proof. This proof is specific for the team and the challenge. Since it is a public proof, the platform and other teams can audit and confirm that the team has indeed solved the challenge, without being able to deduce the flag. We implemented NIZKCTF as an open-source project and its architecture includes software elements such as a Git-based centralized repository (i.e., to commit the flags and receive the event updates), and a continuous integration system (i.e., to automatically merge all requests of teams in the repository of the competition). Any CTF promoter can instantiate and use NIZKCTF in their competition. After validating the usefulness and security of NIZKCTF in two smaller competitions, we adopted the platform to host the 2017 edition of Pwn2Win, an internationally advertised and bilingual Brazilian CTF. During and after the CTFs, teams were not able to compromise the final result.

## History and Background

The first CTF competition apparently was hosted during DEFCON 1996 in Las Vegas. In Brazil, hacking challenges are usually conducted *in loco* during national security events such as the Hackers 2 Hackers Conference ([www.h2hc.com.br](www.h2hc.com.br)) since at least 2007. The first widely advertised jeopardy-style competition was Hacking n' Roll ([www.insert.uece.br/en/events](www.insert.uece.br/en/events)), which began to be organized in 2010 by the Information Security Research Team of the State University of Ceará (UECE). In 2014, more competitions started to appear, such as the Pwn2Win (*on line*) and Hackaflag (*in loco*) CTFs. Despite challenges being enunciated only in Portuguese, a Ukrainian team participated in the 2014 editions of Pwn2Win and Hacking n' Roll, increasing the interest for bilingual and internationally advertised competitions. In March 2016, Pwn2Win became the first Brazilian CTF listed in the CTFtime ([ctftime.org](ctftime.org)) international index, followed by 3DSCTF in December of the same year. Figure 1 shows the growing trend in active participation of Brazilian teams in international competitions.

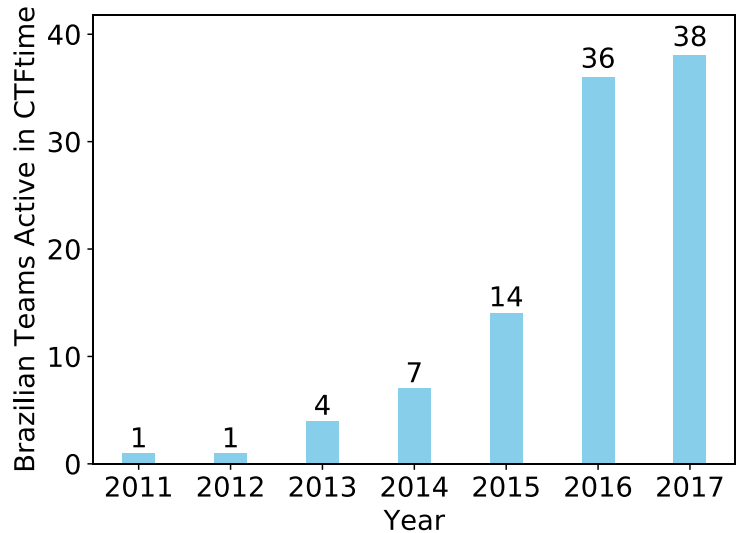

Figure 1. After the first bilingual Brazilian competitions were promoted in 2016, there was an increase in the number of internationally active Brazilian teams indexed by CTFtime.

As organizers of the Pwn2Win CTF, the authors faced scalability issues during the 2016 edition. In the last couple of hours of the competition, the main software platform became nearly unable to render the scoreboard due to the amount of concurrent access. Although this issue could be alleviated by adopting a more efficient open source platform, there was a growing apprehension of facing denial-of-service attacks or even having flags leaked due to a software security flaw or server misconfiguration. As newcomers to the international CTF scene, this could undermine the reputation of Brazilian competitions as a whole, which motivated this work toward a more robust and transparent CTF platform.

## Problem Statement

The impact of software vulnerabilities in CTF platforms is high and gives a disproportionate competitive advantage to skilled teams able to exploit the platform instead of solving the competition challenges. This paper addresses this problem by improving the security of the competition platform and intends to answer the following questions: (i) how to guarantee integrity and a minimum level of fairness in a jeopardy-style CTF competition, preventing teams from stealing flags by exploiting the platform? (ii) how to ensure auditability, allowing anyone to verify whether teams really solved the challenges according to the points presented in the scoreboard? (iii) how to replicate the information required for checking correctness of solutions to the players, to reduce the impact of eventual platform unavailability?

Unfortunately, it is common to find vulnerabilities in CTF platforms, as illustrated by two high-profile recent examples:

- RC3 CTF 2016: For a moment during this CTF, the first place team had 3590 points. Suddenly, a team named "The board is vulnerable, please contact admin@seadog007.me" appeared in the scoreboard with 4500 points (github.com/seadog007/RC3-CTF-2016-scoreboard). Figure 2 shows the record of this fact.
- CODEGATE CTF 2016 Finals: During this CTF, a team discovered that the server hosting one of the challenges had an old kernel version and was vulnerable to OverlayFS privilege escalation (CVE-2016-1576). The team members were able to gain root access and get some "free flags" (although they claim that they did not submit these flags until they really solved the challenges). Tracing the system calls of the SSH server, they waited for an administrator to log in and were able to get their password. Then, the team noticed that other servers (including the scoreboard) had the same password. After visiting the platform servers and having fun, they stopped the intrusion and proceeded to play as usual (mslc.ctf.su/wp/codegategate).

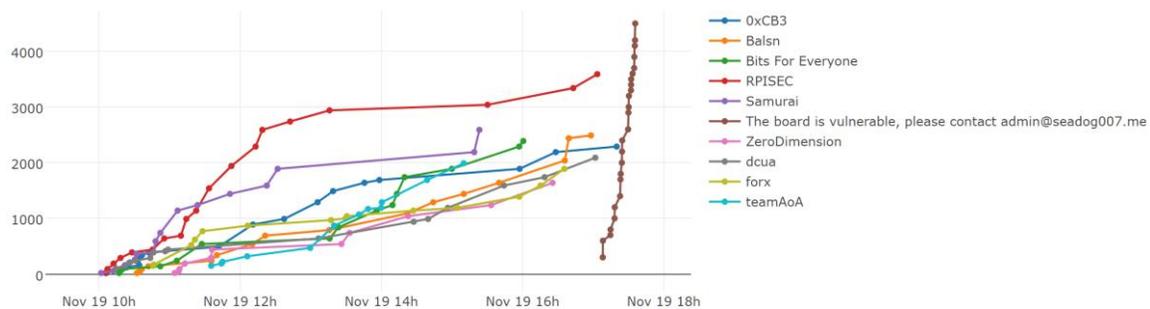

Figure 2. In RC3 CTF 2016, hackers exploited the scoreboard to report the vulnerability to the competition administrators.

## Related Work

Concerns about securing a CTF platform against attacks or just making the competition more fair are not new. In 2015, University of Birmingham and Imperial College London[3] announced a virtual machine (VM) containing vulnerable services and challenges: each student runs the VM locally and attempts to solve challenges inside the VM as they are made available. Solving a challenge reveals a flag that is unique to a particular VM instance, allowing for the detection of collusion between students. As well as acquiring flags, students also had to provide traditional written answers to questions and sit an examination.

In 2015, Carnegie Mellon University[4] released an automatic problem generator (APG) for CTF competitions, where a given challenge is not fixed, but rather can have many different automatically generated problem instances. APG offers players a unique experience and can facilitate deliberate practice where problems vary just enough to make sure a user can replicate the solution idea. APG also allows competition administrators to detect when users submit a copied flag from another user to the scoring server.

There are other works regarding problems that may affect the overall quality of CTFs.[5–8] Chung *et al.*[5] present insights and lessons learned from organizing CSAW, one of the largest and most successful CTFs. Vigna *et al.*[6] present lessons learned in running iCTF, the world's largest attack/defense CTF, in which online players from several countries compete against each other. Chapman *et al.*[7] present the competition design of PicoCTF, as well as an evaluation based on survey responses and website interaction statistics, and insights into the students who played.

Despite the relevance, these works do not address solutions for the security problem of players attacking the CTF platform. A first issue is players attacking the platform to steal the flags. Although common sense dictates that flags should be processed with a one-way function, similarly to how passwords are usually stored, Table 1 shows that most of the previously existing open-source CTF platforms do not support this feature. The only exception is PicoCTF Platform 2, which checks whether flags are correct by calling a script specific to each challenge. The script could be configured to check against a protected flag, although this use case is not a built-in feature or documented at all.

Our proposed solution is discussed next.

*Table 1. Comparison between NIZKCTF and previously existing open source platforms*

| Platform | URL | Supports protected flags | Supports regex flags |
| --- | --- | --- | --- |
| NIZKCTF | github.com/pwn2winctf/nizkctf-tutorial | yes | no |
| CTFd | github.com/CTFd/CTFd | no | yes |
| FBCTF | github.com/facebook/fbctf | no | no |
| HackTheArch | github.com/mcpa-stlouis/hack-the-arch | no | yes |
| Mellivora | github.com/Nakiami/mellivora | no | no |
| NightShade | github.com/UnrealAkama/NightShade | no | yes |
| PicoCTF Platform 2 | github.com/picoCTF/picoCTF-Platform-2 | yes | yes |
| RootTheBox | github.com/moloch--/RootTheBox | no | yes |
| PyChallFactory | github.com/pdautry/py_chall_factory | no | no |

## Competition Protocol

This section describes the novel competition protocol employed by NIZKCTF. A threat model for CTF competitions is defined, followed by a formalization of the main theoretical tools and requirements for auditability, together with textual descriptions for the security properties and employed schemes for clarification.

### Threat Model

In a CTF competition, the adversary is a player or team interested in exploiting vulnerabilities in the platform, instead of solving challenges, to obtain advantages such as stealing flags or manipulating the scoreboard. Due to the difficulty in developing a vulnerability-free system, NIZKCTF relies on cryptographic primitives, such as zero-knowledge proofs, to provide security properties.

Another form of adversarial behavior is a team who wants to submit flags in place of another team without their consent (e.g., to harm a specific team by making it fall behind on the scoreboard). For this reason, NIZKCTF makes a zero-knowledge proof to be unique to a particular team.

The protocol alone does not protect against flag sharing, i.e., teams copying and submitting flags from others. For a solution that addresses this type of adversary, recall the automatic problem generation proposed by Burket *et al.*[4] Since in a competition based on NIZKCTF it is possible to have automatic problem generation, our proposal can be extended to support protection against a flag-sharing adversary.

**Zero-knowledge Proof of Knowledge**

Let $L$ be a NP language such that $pk_i \in L$ iff there exists a witness $sk_i$ yielding $M_L(pk_i, sk_i) = 1$, where $M_L$ is a polynomial-time Turing machine. Let us also assume that the probability of computing $sk_i$ from $pk_i$ in polynomial time is negligible.

A non-interactive zero-knowledge (NIZK) proof of knowledge[9] is a cryptographic scheme through which a prover knowing $sk_i$ can convince a verifier of that fact, satisfying the following properties:

- Non-interactivity: Access to a public set of common parameters and to the contents of the proof itself must be sufficient to verify a proof. Since no interaction with the prover is required, any party interested in acting as a verifier may do so.
- Completeness: If $pk_i \in L$ the proof generated by an honest prover knowing $sk_i$ must be accepted by an honest verifier:
$$\sigma = \text{Prove}(pk_i, sk_i) \Rightarrow \text{Verify}(pk_i, \sigma) = 1$$
- Validity: Any probabilistic polynomial-time (PPT) prover who does not know $sk_i$ must have negligible probability of success $\epsilon$ of convincing a verifier. Equivalently, for every possible PPT prover P (even malicious ones) there exists a knowledge extractor Extract that, given oracle access to P, is able to extract $sk_i$ with overwhelming probability $(1-\epsilon)$ every time P succeeds in completing a new proof:
$$\Pr[\sigma \leftarrow P(pk_i, sk_i); sk_i' \leftarrow \text{Extract}(pk_i, \sigma):$$
$$M_L(pk_i, sk_i') \vee ((pk_i, sk_i) \in Q) \vee$$
$$\neg \text{Verify}(pk_i, \sigma)] = 1 - \epsilon,$$
where $Q$ denotes a query tape which registers all previous queries that have been sent to a prover.
- Zero-knowledge: The proof discloses no information about $sk_i$ besides the fact that the prover knows its value. Equivalently, for every possible PPT verifier V (even malicious ones) there exists a simulator Sim that, given oracle access to V, is able to convince V with a negligible difference in probability $\epsilon$ when compared to an honest prover, even though Sim has no knowledge of $sk_i$:
$$|\Pr[\sigma \leftarrow \text{Prove}(pk_i, sk_i); b \leftarrow V(pk_i, \sigma) : b = 1]$$
$$-\Pr[\sigma \leftarrow \text{Sim}(pk_i); b \leftarrow V(pk_i, \sigma) : b = 1]| = \epsilon.$$

In NIZKCTF, values $pk_i$ are publicly disclosed, but the corresponding $sk_i$ which allows proving $pk_i \in L$ is kept secret. Every player holds a witness $sk_t$ attesting membership to their team $t$. When a player solves a challenge $c$ of the competition, they obtain a witness $sk_c$ asserting they hold the answer to the challenge. In order to earn points for their team, the player needs to publicly prove simultaneous knowledge of $sk_t$ and $sk_c$. The concept of simultaneous knowledge is formalized by performing proofs on an auxiliary NP language $pk_t \parallel pk_c \in L'$ such that $pk_t \parallel pk_c \in L'$ iff there exists a witness $sk_t \parallel sk_c$ such that $M_{L'}(pk_t \parallel pk_c, sk_t \parallel sk_c) = M_L(pk_t, sk_t) \wedge M_L(pk_c, sk_c) = 1$, where the operator $\parallel$ denotes string concatenation.

Different approaches exist for proving knowledge of witness $sk_t \parallel sk_c$, but their practicality depends on the exact choice of $M_L$ and, consequently, of $M_{L'}$. A first approach would be a general-purpose non-interactive zero-knowledge (NIZK) proof system, but the generality comes with a price, in the form of substantial processing time and large public parameters required to construct a proof.

**A scheme based on digital signatures.** The approach proposed and implemented in NIZKCTF consists in choosing a $M_L(pk_i, sk_i)$ which verifies whether $sk_i$ is the private key corresponding to the public key $pk_i$ in a digital signature scheme. This choice allows us to reduce our proof of knowledge problem to that of digitally signing messages, whose implementation is simpler and more efficient than any known general-purpose NIZK proof system.

The Schnorr signature scheme and its key-prefixed variant over elliptic curves EdDSA[10] satisfy completeness, validity and zero-knowledge properties under the assumption that the discrete logarithm problem over elliptic curves (ECDLP) is hard. [11–12] However, incorrectly composing two signatures when constructing a proof of simultaneous knowledge may undermine the validity of such properties.

Let the following be the primitives of a secure digital signature scheme:

$$\text{Sign}(sk_i, m) = s \parallel m \quad (1)$$

$$\text{Open}(pk_i, s \parallel m) = \begin{cases} m, & \text{if } s \text{ is valid} \\ \bot, & \text{otherwise} \end{cases} \quad (2)$$

Equation 1 signs the message $m$ using the private key $sk_i$, then outputs the signature $s$ prepended to the message $m$. Equation 2 verifies whether $s$ is a valid signature for $m$ produced by the private key $sk_i$ corresponding to the public key $pk_i$, then outputs the original message $m$ if the signature is valid, or $\bot$ if it is invalid.

We propose the following scheme to prove knowledge of the witness $sk_t \parallel sk_c$ corresponding to $pk_t \parallel pk_c \in L'$:

$$\text{Prove}(pk_t \parallel pk_c, sk_t \parallel sk_c) = \text{Sign}(sk_c, \text{Sign}(sk_t, c)) \quad (3)$$

$$\text{Verify}(pk_t \parallel pk_c, \sigma) = \begin{cases} 1, & \text{if } m = c \\ 0, & \text{if } m \neq c, \end{cases} \quad (4)$$

$$\text{where} \quad m = \text{Open}(pk_t, \text{Open}(pk_c, \sigma))$$

We argue that Equations 3 and 4 satisfy the properties of a NIZK proof of knowledge scheme:

- Non-interactivity: Since $pk_t$, $pk_c$ and the digital signature scheme parameters are public and known to all parties, the proof $\sigma$ can be verified by Equation 4 without interaction with the prover.
- Completeness: Since $\text{Open}(pk_i, \text{Sign}(sk_i, m)) = m$ for all $i$ and for all $m$ such that $M_L(pk_i, sk_i) = 1$, by simply substituting into Equations 3 and 4:

$$\forall (t, c): pk_t \parallel pk_c \in L',$$

$$\text{Verify}(pk_t \parallel pk_c, \text{Prove}(pk_t \parallel pk_c, sk_t \parallel sk_c)) = 1.$$

- Validity: If the digital signature scheme satisfies validity, there exists a knowledge extractor $\text{Extract}(pk_i, s \parallel m)$ able to extract $sk_i$ from the $\text{Sign}(sk_i, m)$ operation implemented by any (possibly malicious) PPT prover P. Therefore, a knowledge extractor Extract' able to extract $sk_t \parallel sk_c$ from P is constructed as follows:

$$\text{Extract}'(pk_t \parallel pk_c, s_c \parallel s_t \parallel c) =$$

$$\text{Extract}(pk_t, s_t \parallel c) \parallel \text{Extract}(pk_c, s_c \parallel s_t \parallel c).$$

Let $Q$ be the query tape of $\text{Sign}(sk_i, m)$, and $Q'$ be the query tape of $\text{Prove}(pk_t \parallel pk_c, sk_t \parallel sk_c)$. Extract' does not succeed if $(sk_t, c) \in Q$, since P may replay $s_t \parallel c$ from a previous run, causing $\text{Extract}(pk_t, s_t \parallel c)$ to fail in extracting $sk_t$. Similarly, Extract' does not succeed if $(sk_c, s_t \parallel c) \in Q$, since P may replay $s_c \parallel s_t \parallel c$ from a previous run, causing $\text{Extract}(pk_c, s_c \parallel s_t \parallel c)$ to fail in extracting $sk_c$. However, since all messages signed by $sk_t$ reference $c$, $(sk_t, c) \in Q \Leftrightarrow (pk_t \parallel pk_c, sk_t \parallel sk_c) \in Q'$. Similarly, since all messages signed by $sk_c$ reference $t$, $(sk_c, s_t \parallel c) \in Q \Leftrightarrow (pk_t \parallel pk_c, sk_t \parallel sk_c) \in Q'$. The definition of validity allows the knowledge extractor to fail when $(pk_t \parallel pk_c, sk_t \parallel sk_c) \in Q'$, thus existence of Extract' proves that validity is satisfied.
- Zero-knowledge: If the digital signature scheme satisfies zero-knowledge of the private key $sk_i$, there exists a simulator $\text{Sim}(pk_i, m)$ able to convince the validity of the signed message $s \parallel m$ to the $\text{Open}(pk_i, s \parallel m)$ operation implemented by any (possibly malicious) PPT verifier V. Therefore, there exists a simulator Sim' able to convince V of the proof validity:

$$\text{Sim}'(pk_t \parallel pk_c) = \text{Sim}(pk_c, \text{Sim}(pk_t, c)).$$

**Requirements for the challenge witness.** Zero-knowledge of the witness $sk_c$ is useful only as long as $sk_c$ cannot be easily found by exhaustive search. Recall $M_L$ and $pk_c$ are public. Therefore, if $sk_c$ does not have sufficient randomness, an offline brute-force attack has non-negligible chance of success in finding its value.

In some CTFs, the flag $f_c$ for a challenge $c$ consists of a random hexadecimal string large enough (e.g., 256 bits, after decoding) to make a brute-force attack unfeasible. In this case, the flag $f_c$ can be used directly as the seed for a deterministic digital signature key pair generator:

$$(sk_c, pk_c) = \text{KeyPair}(f_c)$$

Many competitions, however, adopt password-like flags, such as "CTF-BR{you_mastered_technique_X}". In this case, a slow to compute password-based key derivation function (PBKDF) can be used along with a public salt value $\varphi_c$ to increase the difficulty of an offline brute-force attack:[13]

$$(sk_c, pk_c) = \text{KeyPair}(\text{PBKDF}(\varphi_c, f_c)) \tag{5}$$

In NIZKCTF, we use Equation 5 as a conservative choice to support both types of flags. We adopt scrypt[14] as a PBKDF and adjust its parameters aiming to make a brute-force attack non-viable within the few hours of the competition.

### Auditability

In order to allow CTFs to be openly audited and independently verified, it is necessary to carry out all operations in a database and satisfy the following requirements:

- History preservation: The database must be able to recover a snapshot of its state after each committed transaction and preserve the logical order of these transactions.
- Immutability: Once a transaction is committed, the database must prevent it from being erased. If an application needs to revert data to a previous state, the only way to perform that operation must be by performing a new transaction.
- Replication: Anyone interested in auditing the competition must be able to retrieve and replicate the entire contents and transaction history of the database.

## Implementation

Different instances of NIZKCTF can be constructed by choosing distinct underlying technologies. We selected components for implementing NIZKCTF with the goal of maximizing the usage of free-of-charge hosted services like GitHub (or GitLab) and Amazon cloud services which provide a permanent free tier (AWS Lambda and SNS).

Our implementation (github.com/pwn2winctf/2017) is composed by the following modules: a distributed storage for sharing data (implemented by a Git repository), a continuous integration script for accepting submissions (implemented by an AWS Lambda function), a command-line interface for interacting with the platform, and a web interface for displaying the list of challenges and the scoreboard.

As can be seen in Figure 3, the distributed storage is used for propagating the necessary data while keeping the full change history. Players then interact with the distributed storage by using the command-line interface, which implements the NIZKCTF protocol. A request to merge the new data with the main repository is created and later evaluated by the continuous integration script, which checks the validity of the modifications, then decides to accept or deny the request. After the changes are merged, the web interface starts using the most recent version of the data.

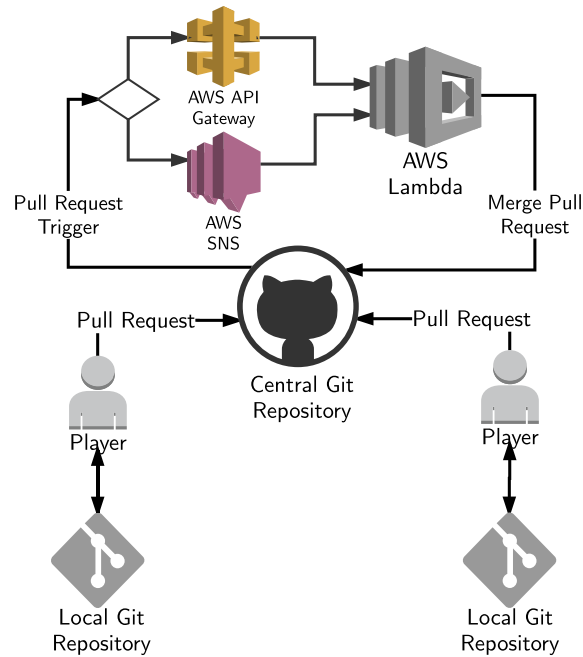

Figure 3. Overview of our implementation. Players modify local Git repositories and create pull requests to the central repository. An AWS Lambda function is triggered to merge the pull request to the central repository if changes are valid.

The distributed storage allows the replication of challenges, team registrations, proofs and other CTF metadata while ensuring that the entire history is preserved. In our implementation, this property is achieved by adopting a Git repository as the database. Git commits are stored as a Directed Acyclic Graph that can be later queried. This allows any participant to audit all changes made, including ordering and timestamps.

When changes are made to the player's local storage, a pull request is created to merge the modifications with the central Git repository managed by the competition's staff. The pull request is evaluated by the continuous integration script and, if accepted, all changed data is incorporated and can be propagated to other teams.

In order to avoid tampering with commit history, the repository is configured to disallow force pushes. Without force pushes, changes committed to the central repository must always descent from the central repository history. In other words, pushed commits are not allowed to modify the commit chain, guaranteeing immutability of previously committed changes.

The continuous integration script is triggered by a GitHub hook to automatically accept pull requests containing team registrations and proof submissions. However, since it does not have access to any privileged information about the challenges, any node with access to the distributed storage can also be used for verifying submissions.

The command-line interface is a Python script used for automating modifications on the distributed storage and uses libsodium for all implementations of cryptography. Currently, the following operations are supported:

- Login: Connects to GitHub or GitLab, generates an API token and creates a fork of the CTF's main repository.
- Register: Registers a new team and creates a pull request.
- Challenges: Lists available challenges with their title, description, categories and rewards.
- Submit: Checks if the challenge's private key can be successfully computed from the flag provided by the competitor, then generates a submission request of the zero-knowledge proof.
- Score: Reads the accepted submissions file and presents the scoreboard.

The web interface uses the repository as a GitHub page which exposes files with an HTTP server. This allows the challenges and scoreboard to be viewed in a more user-friendly way. The interface is implemented using only client-side programming languages (HTML, CSS and JavaScript). Challenges and scoreboard files are loaded using Ajax to give a dynamic feel.

It is worth noting that the distributed storage could also be implemented using a blockchain. This would allow a fully distributed implementation of NIZKCTF in which submission proofs are appended to the blockchain and validated by contest participants. We do not choose such an implementation because it would require a small transaction fee for each submitted proof and extra complexity.

## Validation

To validate our proposal, we first conducted two small CTFs as pilot tests. The first one was Pwn2Win Platform Test Edition – PTE (github.com/pwn2winctf/PTE), a competition for 10 international invited teams. The objective of PTE was to assess the usefulness and security of NIZKCTF. In order to achieve that, we used the Goal, Question, Metric (GQM) paradigm,[15] a mechanism for defining and evaluating goals using measurement.

GQM defines a measurement model on three levels: conceptual level (Goal), operational level (Question) and quantitative level (Metric). GQM templates are a structured way of specifying goals and contains the following fields: purpose, object of study, focus, stakeholder and context. Here is a GQM template to express the goal of our study: The purpose of this study was to *evaluate the usefulness* of *NIZKCTF* when *being used* by the *participants* in a *CTF competition*.

To characterize the measurement object, we defined the research question *RQ1*: Is our implementation of NIZKCTF able to provide the features (e.g., challenges, submissions and scoreboard) of a common CTF?

Pwn2Win PTE had 7 challenges and duration of 12 hours. There were one challenge for exploitation, cryptography, web, networking and miscellaneous, and two about reverse engineering. Teams were able to solve the challenges and there was a scoreboard, just like a common CTF. Players also gave good feedback and no one had objections in using NIZKCTF in future CTFs. Therefore, we support a positive answer for *RQ1*.

From the 10 invited teams, 5 of them scored (solved at least one challenge). Since the CTF had many low-complexity challenges and the invited teams were very experienced (for example, one of the teams was the 2016 second place at CTFtime), we assume that teams that did not score were focused on trying to exploit the platform, as we present next.

Here is another GQM template to express another goal of our study: The purpose of this study was to *evaluate the security* of *NIZKCTF* against *attacks to the platform* from the *participants* in a *CTF competition*.

To characterize the measurement object, we defined the research question *RQ2*: Is any participant able to attack the platform and compromise the CTF result?

To answer this question, we made a bug bounty program, with 450 BRL in cash prizes for teams who found vulnerabilities to compromise the result. During and after (we kept the platform online for 20 days) the CTF, teams were not able to do that. Therefore, we support a negative answer for *RQ2*.

After that, we conducted the second pilot test: SCMPv8 CTF ([github.com/scmp-ctf/SCMPv8](github.com/scmp-ctf/SCMPv8)), during the 8th Computer Science and Engineering Week (SeComp) at the Federal University of São Carlos (UFSCar). The purpose of this study was to *evaluate the ease of use* of *NIZKCTF* when *being used* by *Brazilian undergraduate students* in a *CTF competition*.

To characterize the measurement object, we defined the research question *RQ3*: Would students take their time to learn how to use NIZKCTF when presented with an incentive?

We conducted SCMPv8 in two different tracks: one of them using a traditional web-browser based CTF platform, and the other one using NIZKCTF. Challenges available in the traditional platform were a subset of the ones in NIZKCTF, and players were allowed to participate in both tracks. Winners of the NIZKCTF track received 4 tickets to attend Hackers 2 Hackers Conference (worth a total of 1200 BRL).

Students from other universities were invited to participate remotely. The traditional track received 27 team subscriptions, from which 14 teams solved at least one challenge, and 3 teams eventually solved every challenge available. The NIZKCTF track received 5 team subscriptions, from which 4 teams solved at least one challenge, and 2 teams eventually solved every challenge available. From this data, we infer a mixed answer for *RQ3*, since only the most skilled teams took their time to use NIZKCTF.

## Deployment

Pwn2Win 2017 was the first large-scale CTF on which NIZKCTF was deployed in production. During the 48 hours of the competition, 283 teams registered, with 207 scoring at least one point. From those teams, 30 were from Brazil and 21 scored points. The best placed Brazilian teams finished at the 11th, 27th and 30th ranking positions. The main repository used to keep track of the submissions had 1372 clones and more than 3800 commits submitted by more than 320 players (repository contributors). The full record of submissions can be found in the submissions repository ([github.com/pwn2winctf/2017submissions](github.com/pwn2winctf/2017submissions)).

During the contest, the scalability of this model was tested and validated. The page containing the challenge listing and scoreboard had 2.7 million HTTP requests. The flag submission process also worked as expected during different loads. The mean time between submissions was 3 minutes, reaching one submission every 2 seconds during peak hours. Figure 4 presents the number of flag submissions by hour.

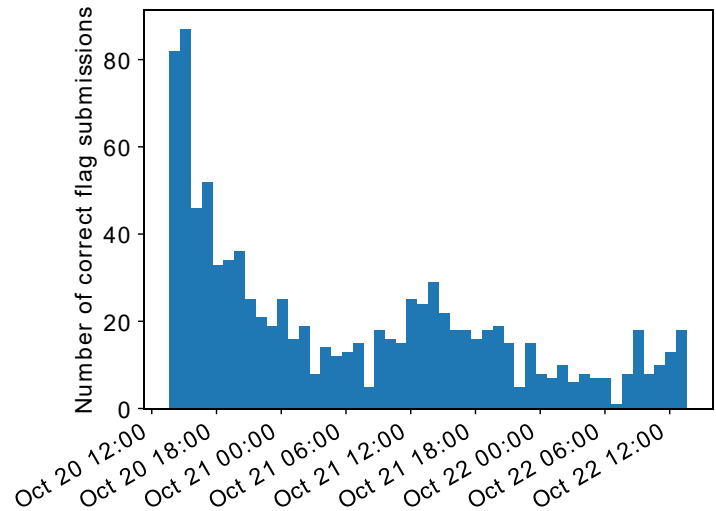

Figure 4. The Pwn2Win CTF 2017 platform processed at least 5 correct flag submissions per hour, except for a brief hiatus in October 22 at 6:00 UTC-2.

In moments of many concurrent submissions, we started observing failures in GitHub's merge API. This was an undocumented behavior that was not triggered in small-scale tests. To mitigate this problem, we patched the application with a hotfix to retry the API request in case of errors. This solution was enough to ensure the correct behavior of the system.

With the growing interest in CTFs, there is a need for a secure and auditable platform. We presented a novel platform called NIZKCTF: Non-Interactive Zero-Knowledge Capture the Flag. Through the security properties assured by the cryptographic mechanisms, we claim that a CTF running on NIZKCTF is more secure than when running a traditional platform.

The implementation of NIZKCTF was tested in three different jeopardy-style CTFs with different characteristics and sizes. In these events, the proposed system was shown to be a safe, scalable and openly auditable alternative to current CTF platforms. During this period no team was able to take advantage of the platform, no tampering happened, and the full history was available to every team.

As future work, we intend to continue using our proposed platform in upcoming competitions. We also intend to develop a fully web-based client using Ajax to commit directly through GitHub/GitLab API endpoints. This will reduce the barriers for team registration and participation without compromising the benefits of the platform. Extending the ideas presented here to attack/defense CTFs would also be an interesting exercise. We hope that NIZKCTF will further engage the Brazilian community in CTF competitions and cybersecurity training.

**Paulo Matias** is an Assistant Professor in the Department of Computing at the Federal University of São Carlos. Contact at matias@ufscar.br.

**Pedro Barbosa** holds a PhD degree in Computer Science from the Federal University of Campina Grande. He works as a security engineer for a reputable multinational company. Contact at pedroyossis@copin.ufcg.edu.br.

**Thiago N. C. Cardoso** is the CTO of Hekima, a Brazilian startup based in Belo Horizonte which develops machine learning solutions. Contact at thiago.cardoso@hekima.com.

**Diego M. Campos** is a PhD candidate in Electrical Engineering at the University of São Paulo. He is the information security coordinator of a reputable Brazilian multinational bank. Contact at diegomcampos@usp.br.

**Diego F. Aranha** is an Assistant Professor in the Institute of Computing at the University of Campinas. Contact at dfaranha@ic.unicamp.br.